\documentclass[twocolumn,amssymb,amsmath,superscriptaddress,aps,pre,reprint]{revtex4-1}
\usepackage{graphicx}
\usepackage{dcolumn}
\usepackage{bm}
\usepackage{subfigure}

\begin{document}

\title{Persistence and failure of mean-field approximations adapted to a class of systems of delay-coupled excitable units}

\author{Igor Franovi\'c}
\affiliation{Faculty of Physics, University of Belgrade, PO Box 44, 11001 Belgrade, Serbia}

\author{Kristina Todorovi\'c}
\affiliation{Department of Physics and Mathematics,Faculty of Pharmacy, University of Belgrade,
Vojvode Stepe 450, Belgrade, Serbia}

\author{Neboj\v{s}a Vasovi\'c}
\affiliation{Department of Applied Mathematics, Faculty of Mining and Geology, University of Belgrade, PO Box 162, Belgrade, Serbia}

\author{Nikola Buri\'c}
\email{buric@ipb.ac.rs}
\affiliation{Scientific Computing Lab., Institute of Physics, University of Beograd, PO Box 68, 11080 Beograd-Zemun, Serbia}%

\date{\today}

\begin{abstract}
We consider the approximations behind the typical mean-field model derived for a class of systems made up of type II excitable units influenced by noise and coupling delays. The formulation of the two approximations, referred to as the Gaussian and the quasi-independence approximation, as well as the fashion in which their validity is verified, are adapted to reflect the essential properties of the underlying system. It is demonstrated that the failure of the mean-field model associated with the breakdown of the quasi-independence approximation can be predicted by the noise-induced bistability in the dynamics of the mean-field system. As for the Gaussian approximation, its violation is related to the increase of noise intensity, but the actual condition for failure can be cast in qualitative, rather than quantitative terms. We also discuss how the fulfilment of the mean-field approximations affects the statistics of the first return times for the local and global variables, further exploring the link between the fulfilment of the quasi-independence approximation and certain forms of synchronization between the individual units.
\end{abstract}


\maketitle

When modeling macroscopic systems comprised of coupled oscillating units, one is often required to incorporate noise and interaction delays, whose separate or combined effects may substantially alter the "bare" dynamics, unattended by these two ingredients. The systems where coaction of noise and delays should be taken into account appear to be common, rather than rare \cite{TP01,PJ08}, with the most prominent examples derived from the biophysiological context \cite{CDK97} or involving laser dynamics \cite{M01}. The interplay of noise and delays becomes especially intricate if the units making up the system are not self-oscillating, but excitable \cite{I07}. For such a setup, the local and collective dynamics rest on the competition between the delay-driven and the noise-induced oscillation modes \cite{prl2,prl1}.

In mathematical terms, the described models are usually stated in terms of systems of nonlinear stochastic delay-differential equations (SDDE), whose general form is given by
\begin{align}
d{\mathbf x}_i(t)=f({\mathbf x}_i(t))+\sum_{i,j}^Ng_{ij}({\mathbf x}_i(t),{\mathbf x}_j(t-\tau_j))
+\sigma_idW_i, \label{eq1}
\end{align}
where $i,j=1,...N$, ${\mathbf x}_i$ are vectors of dynamical variables for the i-th unit, $f$ is a nonlinear function, $\tau_i$ are the coupling delays and $dW_i$ are stochastic increments of the independent Wiener
processes. For systems like \eqref{eq1}, the standard Fokker-Planck formalism can rarely provide useful
results that may serve for qualitative analysis of stochastic stability and stochastic bifurcations \cite{analytic1,analytic2,analytic3,analytic4}, with its use severely constrained by the non-Markovian
character and nonlinearity of the equations \cite{GHL99}. In particular, for univariate systems one may only consider the limits of small delay or delay larger than the system's correlation time \cite{GHL99,GLV09,GBV11,BVTH05}, whereas for the setup involving multiple units, even the delay-free case under the assumption of molecular chaos cannot be analytically resolved \cite{ZSSN05}, though efficient numerical methods are available \cite{ABR04}.

The failure of Fokker-Planck formalism implies the necessity for considering approximate methods, one of them being the mean-field (MF) approach. The gains from the latter can in general be cast as twofold. On one hand, an accurate MF model substantially reduces the computational time for numerical integration \cite{numerical}, which for the exact system grows as $\sim N^2$ with its size \cite{Hasegawa2}. The other gain lies in the ability to translate the problem of stochastic bifurcations displayed by the exact system into bifurcations of the compact deterministic MF system \cite{MFA1,MFA2,MFA3}. The stochastic bifurcations have so far been characterized phenomenologically, relying on the point that a certain time-averaged quantity, such as the asymptotic probability distributions of the relevant variables or the associated power spectra undergo some qualitative change \cite{H04,MFA1,ABR04,A99}. For instance, it has been shown that the stochastic Hopf bifurcation from a stochastically stable fixed point to a stochastically stable limit cycle is accompanied by the loss of Gaussian property for the asymptotic distributions of the appropriate variables \cite{MFA1}. In parallel, the degree by which the distribution departs from the normal one with supercriticality depends on the particular system at hand. However, the onset of stochastic bifurcation and the loss of Gaussian property for asymptotic distributions alone do not imply the failure of the MF approximate model. In fact, such a correspondence would apply only if the MF approximation were based on the notion that the described stochastic process is a Gaussian one. Nevertheless, such a requirement is too strong, in a sense that the validity of the MF model is then satisfied trivially. Moreover, examples have already been found where the MF system can accurately predict the properties of the oscillatory state, including the oscillation frequency \cite{derivation1,derivation2}.

The arguments above justify a more elaborate research on the approximations behind the MF model and the conditions for their validity. On one hand, one may ask whether the MF approximations (MFAs) are universal as might be expected at first sight, or should in fact be modified to account for the properties of the system at hand. The second question we address is whether the dynamics of the MF model itself may point to parameter domains where the MF approximation fails. At variance with the earlier work, this set of issues is completely unrelated to the asymptotic distributions of the underlying variables.

As an example of a system conforming to \eqref{eq1}, we consider an assembly of delay-coupled noisy class II excitable units \cite{I07}, represented by the generic Fitzhugh-Nagumo model. For such a system, we demonstrate that the MFAs should be precisely adapted to its properties, explicitly incorporating the key ingredients of type II excitability, such as relaxation character of oscillations, in the definition of the MFAs and the methods by which their validity is verified. The main benefit from the refined definitions is that they provide rationale on why the predictions provided by the MF model successfully extend to parameter domains admitting oscillatory states, where the trivial Gaussian approximation would necessarily be violated. The other important point shown in the paper is that the dynamics of the MF model may indicate in the self-consistent fashion the domains where one of the MFAs we introduced fails. In particular, such a breakdown of the MF approximate model is found to be linked with the noise-induced bistability in the MF dynamics. Note that the term noise-induced bistability refers to either coexistence of the fixed point and a limit cycle or two limit cycles, which emerge due to action of the noise intensity parameter. The appearance of these regimes is associated with the global fold-cycle (tangent) bifurcation controlled by the noise intensity.

The paper is organized as follows. In section \ref{MFAs}, we precisely define the two main approximations behind the MF model, dubbed Gaussian approximation and the quasi-independence approximation, whose formulations are adapted to reflect the qualitative properties of type II excitable units. Section \ref{tests} is concerned with identifying the typical dynamical scenarios where the MFAs are seen to hold or fail, whereby the tests applied to verify the validity of MFAs are accommodated to class II excitable systems. In section \ref{bistab} we discuss one of the key points, consisting in the ability to deduce the parameter domains where the quasi-independence approximation fails solely by the dynamics of the MF model. Section \ref{misc} deals with several miscellaneous topics, including how the fulfilment of MFAs is manifested in the statistics of the first return times for the local and collective variables, as well as the link between synchronization and the fulfilment of the quasi-independence approximation. Section \ref{conc} contains a summary of the main points of the paper.

\vspace{-0.5cm}
\section{Exact system, MFAs and MF model} \label{MFAs}
\subsection{Background on the exact system} \label{back}

Validity of MFAs is analyzed in case of a collection of N Fitzhugh-Nagumo excitable units, whose dynamics is set by:
\vspace{-0.3cm}
\begin{align}
\epsilon dx_i&=(x_i-x_i^3/3-y_i)dt+{c\over N}\sum_{j=1}^N (x_j(t-\tau)-x_i)dt\nonumber\\
dy_i&=(x_i+b)dt+\sqrt{2D}dW_i, i=1,\dots N \label{eq2}
\end{align}
Each unit interacts with every other via diffusive delayed couplings, whereby the coupling strength $c$ and the time-lag $\tau$ are taken uniform. Parameters $\epsilon=0.01$ and $b=1.05$ are such that the isolated units display excitable behavior, having stable fixed point (FP) as the only attractor. The terms $\sqrt{2D}dW_i$ represent stochastic increments of the independent Wiener processes, viz. $dW_i$ satisfy $E(dW_i)=0$, $E(dW_idW_j)=\delta_{i,j}dt$, where $E()$ denotes the expectation over different realizations of the stochastic process.

Having proposed that the nontrivial conditions for the fulfilment of the MFAs derive from the qualitative properties of the underlying dynamics, we first summarize the typical regimes exhibited by $(x_i(t),y_i(t))$, beginning with the deterministic case $D=0$. For small $c$ and $\tau$, the only attractor of each unit is FP and the dynamics is excitable. For larger $c$ and/or larger $\tau$, the FP undergoes a Hopf bifurcation and the asymptotic dynamics resides on a stable limit cycle (LC). The LC conforms to relaxation oscillations, with two clearly distinguished  slow branches, the refractory and the spiking one, and two fast transients in between, cf. Fig. \ref{Fig1}(b) where small noise perturbations are added. Small D induces small fluctuations around the attractor of the deterministic dynamics. If the latter motion lies on LC, the impact of $D$ is reflected mostly in the fluctuations of phase of the oscillatory dynamics between the different stochastic realizations. Apart from the increase of fluctuation amplitudes, enhancing $D$ may give rise to the transition from the stochastically stable FP to the noise induced spiking. The latter can appear as nearly periodic or irregular depending on $c, \tau$ and $D$. It is known that in systems of excitable units subjected to $D$ and $\tau$, the length of inter-spike intervals (ISIs) is influenced by the competition between two characteristic time scales \cite{prl1,prl2}. One is set by the self-oscillation "period" $T_0(D)$ obtained for $\tau=0$, whereas the other is adjusted with $\tau$. Loosely speaking, for $\tau<T_0(D)$ and intermediate $c$, the noise-led dynamics characterized by $T_0(D)$ prevails over the delay-driven one unless $\tau$ is commensurate or comparable to $T_0(D)$. This paradigm may carry over to the collective motion due to synchronization of individual units.

\vspace{-0.4cm}
\subsection{Formulation of MFAs} \label{formulation}

The first MFA derives from the strong law of large numbers, by which the sample average
$S_N=N^{-1}\sum\limits_{i=1}^{N}s_i$ of $N$ independent and identically distributed random variables $s_i$ converges almost surely to the expectation value $E(s_i)$ for $N\rightarrow\infty$. How $S_N$ approaches $E(s_i)$ for large, but finite $N$ and finite variances of $s_i$ distributions $\sigma^2$, is specified by the central limit theorem, which implies that $S_N$ follow the normal distribution $\mathcal{N}(E(s_i),\sigma^2/N)$. In our setup,
the subsets $\{x_i(t)|i=1,\dots N\}$ and $\{y_i(t)|i=1,\dots N\}$ at any given $t$ are obviously not made up of independent variables, but one may still consider the influence of interaction terms negligible if $N$ is sufficiently large. The latter is referred to as the quasi-independent approximation (QIA), whose precise formulation is: \\
\vspace{0.1cm}
{\textbf{Approximation $1$. (QIA)} \it random variables $\{x_i(t)|i=1,\dots N\}$ and $\{y_i(t)|i=1,\dots N\}$ for each $t$ and sufficiently large N satisfy the approximate equalities}:
\vspace{-0.3cm}
\begin{align}
X(t)&\equiv\frac{1}{N}\sum_i^N x_i(t)\approx E(x_i(t))\nonumber\\
Y(t)&\equiv\frac{1}{N}\sum_i^N y_i(t)\approx E(y_i(t)) \label{eq3}
\end{align}
On the left of (\ref{eq3}) are the {\it spatial averages}, used to define the global variables $X(t)$ and $Y(t)$.
Note that the method implemented in section \ref{tests} to test the validity of QIA will reflect the relaxation character of oscillations typical for class II excitable systems.

The need for the second approximation becomes apparent after carrying out the spatial average and applying the QIA on (\ref{eq2}). The fashion in which the terms $E(x_i^3(t))$ are to be treated is resolved by the {\it Gaussian approximation} (GA), given as: \\
\vspace{0.2cm}
{\textbf{Approximation $2$. (GA)} \it for most time instances $t_0$, the small random increments $dx_i(t),dy_i(t)$ for  $t\in (t_0,t_0+\delta t)$ can be computed with sufficiently good accuracy by assuming that the random variables $x_i(t),y_i(t)$  for each $i=1,\dots N$ and for $t\in (t_0,t_0+\delta t)$ are normally distributed around $(E(x_i(t)),E(y_i(t)))\approx (X(t),Y(t))$.}

GA is intentionally stated in a weak sense, containing phrases "sufficiently good accuracy" and "for most time instances". The former implies that the approximate solution should have the same qualitative features as the exact one. Nevertheless, the phrase "for most time instances" is crucial, because it specifically targets the class II excitable systems, being introduced to account for the relaxation character of oscillations, as explicitly demonstrated in section \ref{tests}. Further note that the GA does not require $\{x_i(t)|i=1,\dots N\}$ and $\{y_i(t)|i=1,\dots N\}$ to be Gaussian processes over asymptotically large time intervals, but rather to be Gaussian over small intervals $(t,t+\delta t)$, with the latter supposed to hold for most values of $t$. For such $t$ one can express all the higher order moments that appear in the expressions for $dX(t)$ and $dY(t)$ in terms of only the means, viz. $X(t)$ and $Y(t)$, and the second-order moments, including variances $s_x(t)=E(n^2_{x_i}(t)),s_y(t)=E(n^2_{y_i}(t))$ and the covariance $u(t)=E(n_{x_i}(t)n_{y_i}(t))$, where $n_{x_i}(t)=X(t)-x_i(t), n_{y_i}(t)=Y(t)-y_i(t)$. Here, the QIA is reflected in the fashion in which the spatial and the stochastic averages are related. Use of GA in deriving the MF model rests on the notion that the fraction of time where GA fails will not introduce significant differences between the MF and the exact solutions.

The MF counterpart of (\ref{eq2}), incorporating QIA and GA, has been derived in \cite{derivation1}. It constitutes the following system of five deterministic DDE:
\begin{align}
\epsilon{\dot{m_x}}&=m_x(t)-m_x(t)^3/3-s_x(t)m_x(t)-m_y(t)\nonumber\\
&+c(m_x(t-\tau)-m_x(t)),\nonumber\\
{\dot{m_y}}&=m_x(t)+b,\nonumber\\
{\epsilon\over2}{\dot{s_x}}&=s_x(t)(1-m_x(t)^2-s_x(t)-c)-u(t)\nonumber\\
{1\over2}{\dot{s_y}}&=u(t)+D,\nonumber\\
{\dot{u}}&=(u(t)/\epsilon)(1-m_x(t)^2-s_x(t)-c)-s_y(t)/\epsilon+s_x(t) \label{eq6}
\end{align}
assuming that the MF solutions satisfy $m_x(t)\approx X(t), m_y(t)\approx Y(t)$. Note that some more sophisticated MF approaches \cite{Hasegawa1,Hasegawa2} adopt the Gaussian decoupling approximation, yet do not require QIA, as their final form accounts for spatial averages of fluctuations of both local and global variables.

\vspace{-0.3cm}
\section{Testing the validity of MFAs and the generic regimes where they hold or fail} \label{tests}

The goal is to first explain the two generic scenarios where both the MFAs hold, outlining the parameter domains where the pertaining local and global dynamics typically occur. We also illustrate the case where both the MFAs fail, independently demonstrating that the GA and the QIA are violated. As indicated in the Introduction, the methods applied to verify the validity of the MFAs for the oscillatory state are adapted to the essential properties of the class II excitable systems. Note that in this section, which contains the numerical results on the exact system, one is primarily concerned with the fulfilment of GA. This is done deliberately, given that the breakdown of QIA can be deduced from the dynamics of the MF model, as demonstrated in section \ref{bistab}.

Intuitively, one would expect that both the MFAs are satisfied if $c$ and $D$ are small. Though this is indeed so, the general conclusion on simultaneous validity of both the MFAs is less straightforward, and should refer to the qualitative properties of the system's dynamics, rather than alluding to certain parameter domains. As a preview of this result, it may be stated that the GA and the QIA hold if the local and the global dynamics are characterized by a single attractor of the same type, either a FP or a LC, provided that $D$ is not too large. Conversely, if the local and the collective variables yield qualitatively different dynamics or exhibit multistability, the validity of either or both the MFAs is lost. Nevertheless, note that the separate conditions for failure of each of the MFAs can be put in a more succinct form, which will be clarified below.

\begin{figure*}[t]
\centering
\includegraphics[scale=0.275]{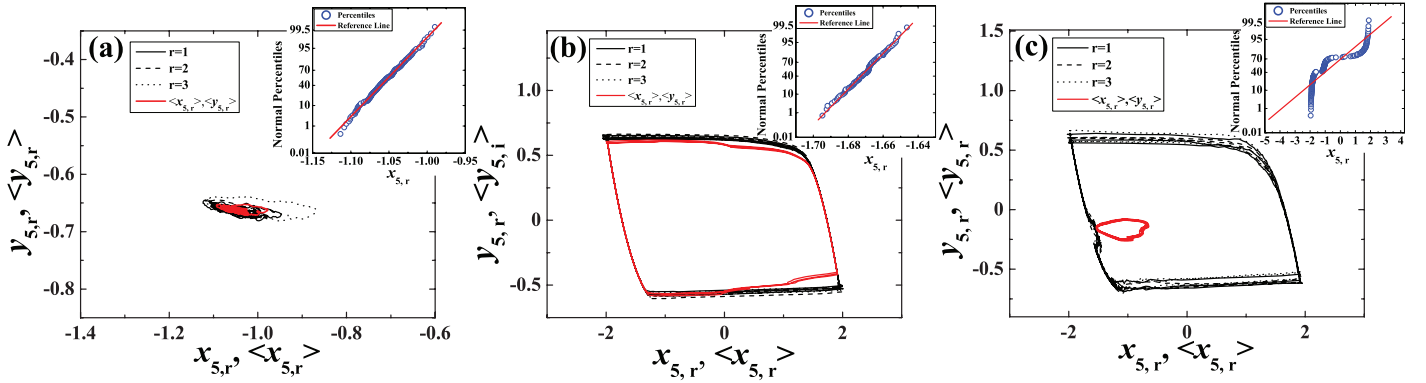}
\caption{(Color online) Impact of noise on validity of GA. (a) and (b) refer to typical scenarios where the GA holds, while (c) concerns one of the cases where it fails. For an arbitrary unit, here denoted with index $5$, the main frames of (a), (b) and (c) show orbits corresponding to three different stochastic realizations $(x_{5,r},y_{5,r})$ (black solid, dashed and dotted lines), as well as the respective expectations $(E(x_{5,r}),E(y_{5,r}))$ for an ensemble of $10$ realizations, having indicated the latter by the solid red (gray) lines. In the insets of (a), (b) and (c) are displayed the appropriate graphic normality tests, such that the red (gray) lines conform to theoretical data sets obeying the Gaussian distribution, whereas the circles reflect the data for $x_{5,r}$ actually collected over different realizations. The parameter sets are $(c,D,\tau)=(0.1,0.0002,0.2)$ in (a),
$(c,D,\tau)=(0.1,0.0002,2.7)$ in (b) and $(c,D,\tau)=(0.1,0.003,1.5)$ in (c).\label{Fig1}}
\end{figure*}

The discussion above implies that there are two paradigmatic scenarios where both the GA and the QIA hold. By one,
the local and the collective dynamics display stochastically stable FP, whereas in the other, the local and the collective dynamics exhibit the stochastically stable LC. These two cases are addressed in Fig. \ref{Fig1}(a) and Fig. \ref{Fig1}(b), respectively, whereby the intention is to first verify the validity of GA. Before proceeding in this direction, note that the value $c=0.1$, fixed in both instances, is chosen from an intermediate range to stress that the MFAs' validity extends into such a domain. Nonetheless, the data in Fig. \ref{Fig1}(a) are obtained for small $D_1=0.0002$ and small $\tau_1=0.2$, while the setup in Fig. \ref{Fig1}(b) involves $D_2=D_1$ but the much larger $\tau_2=2.7$.

As an illustration on the qualitative similarity between the individual realizations and the expectations, the main frames of Fig. \ref{Fig1}(a) and Fig. \ref{Fig1}(b) show three different stochastic realizations $(x_{5,r},y_{5,r})$, encoded in black solid, dashed and dotted lines, as well as the expectation values $(E(x_{5,r}),E(y_{5,r}))$ over an ensemble of $10$ realizations, having plotted them by the solid red (gray) lines. The data are representative for the dynamics of an arbitrary neuron, whereby the particular example refers to the unit $i=5$. The index $r$ accounts for the realizations. In case of Fig. \ref{Fig1}(a), for any $t$, the expectation closely matches either of the realizations trivially. Nevertheless, for the scenario with the LC attractor, the analogous statement holds if $t$ is such that $(E(x_{i,r}(t)),E(y_{i,r}(t))\approx(X(t),Y(t))$ lies on the slow branches of the given orbit. At variance with this, if $(E(x_{i,r}(t)),E(y_{i,r}(t))$ falls onto one of the transients, the expectation departs substantially from the realizations, cf. Fig. \ref{Fig1}(b). The two latter points are consistent with the relaxation character of oscillations, and as such are accounted for by the definition of GA. Invoking the definition, it follows that the GA's validity is upheld if the number of instances where the expectations closely match the individual realizations strongly prevails over the number of instances where such a correspondence is lost. In other words, the GA holds if the expectations preserve the relaxation character of oscillations exhibited by the realizations. Though this requirement is qualitative in nature, one may still attempt to attribute it a certain quantitative measure. For instance, for the $(c,D_2,\tau_2)$ parameter set from Fig. \ref{Fig1}(b), it may be shown that the ratio of points lying on the transients vs those on the slow branches is small ($n_t/n_s\approx0.1$) over the sufficiently long time period along the trajectory of $(E(x_{i,r}(t)),E(y_{i,r}(t))$.

Figure \ref{Fig1}(c) refers to the case where GA no longer holds. The illustrated local dynamics is obtained for comparably large $D_3=0.003$, $c=0.1$ and intermediate $\tau_3=1.5$, such that the individual stochastic realizations fluctuate around the single LC. However, the noise-induced fluctuations are large enough to throw the different realizations out of step, resulting in the strong misalignment between the pertaining oscillation phases. Therefore, at variance with Fig. \ref{Fig1}(b), the expectation substantially departs from each of the realizations at \emph{any} $t$. For $(c,D_3,\tau_3)$, one can no longer interpret $(E(x_{i,r}(t)),E(y_{i,r}(t))$ in terms of clearly discernible slow and fast motions, so that the ratio $n_t/n_s$ cannot be determined. 

Apart from characterizing it by the $n_t/n_s$ ratio, the GA validity has been tested directly for an arbitrary neuron at the given $(c,D,\tau)$. Having run many different realizations of the processes $x_i(t),y_i(t)$ for the same initial function, we have examined the properties of the distribution of different realizations
$x_{i,r}(t_0+\delta t),y_{i,r}(t_0+\delta t)$ for small $\delta t$, taken to be of the order, in tens or hundreds of iteration steps. For the LC dynamics, $(x_{i,r}(t_0),y_{i,r}(t_0))$ has been set on the refractory branch. The insets of (a), (b) and (c) display graphic normality tests, where the red lines indicate the theoretical percent of data points that would lie below the given value if obeying the Gaussian distribution, while the blue circles refer to the cumulative distribution of $(x_{5,r},y_{5,r})$ for an ensemble of over $200$ realizations. Apparently, the distributions corresponding to $(c,D,\tau)$ in Fig. \ref{Fig1}(a) and \ref{Fig1}(b) are Gaussian, whereas the one for Fig. \ref{Fig1}(c) is not.  Results of the graphic tests are corroborated by the standard numerical Shapiro-Wilk method.

\begin{figure}[t]
\vspace{-0.6cm}
\centering
\includegraphics[scale=0.37]{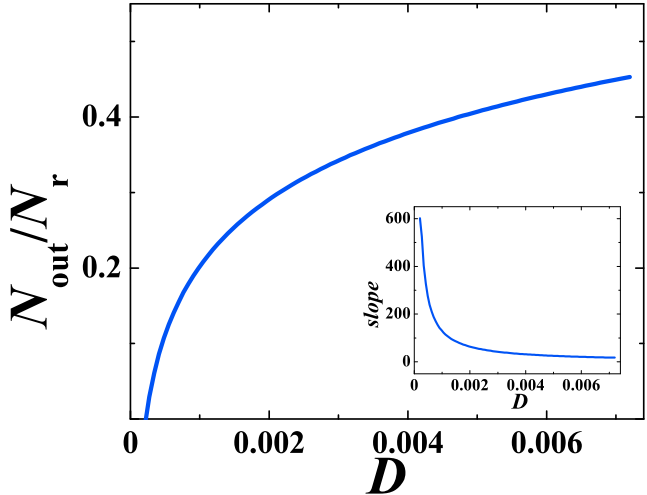}
\caption{(Color online) Estimating rate by which the validity of GA deteriorates with increasing $D$. The plot shows a fraction of stochastic realizations $N_{out}/N_r$ in dependence of $D$ for $(c,\tau)=(0.1,2.7)$. $N_{out}$ accounts for the instances where the representative point escapes the refractory branch of slow motion within the given number of iteration steps $T_{max}$. The results converge to the displayed curve as $N_r$ is increased. The inset refers to the variation of the slope of the curve from the main frame with noise.
\vspace{-0.4cm}
\label{Fig2}}
\end{figure}

Having seen that the criterium for validity of GA is primarily qualitative, it is still of interest to find some indication on how the fulfilment of GA deteriorates with variation of the system parameters. Naturally, the most relevant question is to assess the rate at which the validity reduces with increasing $D$ for fixed $c$ and $\tau$. The quantity appropriate to characterize this is determined as follows. For very small $D=0.0002$, we select an arbitrary neuron and fix a point on the refractory branch of its LC orbit. Then, a large number of different stochastic realizations $N_r$ for the given parameter set $(D,c,\tau)=(0.0002,0.1,2.7)$ is run. The goal is to find the maximal number of iteration steps $T_{max}$, for which the representative point in all the realizations still lies on the refractory branch. Enhancing $D$ while $T_{max}$ is kept fixed, one naturally encounters realizations where the latter condition is no longer satisfied. In Fig. \ref{Fig2} we demonstrate how the fraction of realizations $N_{out}/N_r$ in which the representative point has escaped the refractory branch in less than or exactly $T_{max}$ steps increases with $D$. Along with allowing one to quantify the gradual loss of GA's validity, this dependence may also be interpreted as an indication on how $D$ gives rise to the number of moments $t$ where $(E(x_{i,r}(t)),E(y_{i,r}(t))$ belong to fast transients, rather than the two slow branches. In this context, it is interesting to explain why the curve's slope shows a significant change in behavior around $D_0\approx0.0014$, cf. the inset in Fig. \ref{Fig2}. Below $D_0$, the fluctuations of phase between the different stochastic realizations systematically grow, but the physical picture by which the LC for the expectations $(E(x_{i,r}(t)),E(y_{i,r}(t))$ is described in terms of two pieces of slow motion connected by the two rapid jumps still applies. Nevertheless, about $D\simeq D_0$, such a picture has to be abandoned, because the LC generated by the expectations no longer matches the phase portrait of individual realizations. In particular, the $(E(x_{i,r}(t)),E(y_{i,r}(t))$ cycle lies inside the one for the single realizations, as it fails to reach the latter's spiking branch. Once the framework involving qualitative equivalence between the dynamics of realizations and the expectations has been broken, $N_{out}/N_r$ for $D>D_0$ loses its original meaning, but its steady increase reflects the tendency for growing irregularity in the unit's behavior.

\begin{figure*}[t]
\centering
\includegraphics[scale=0.29]{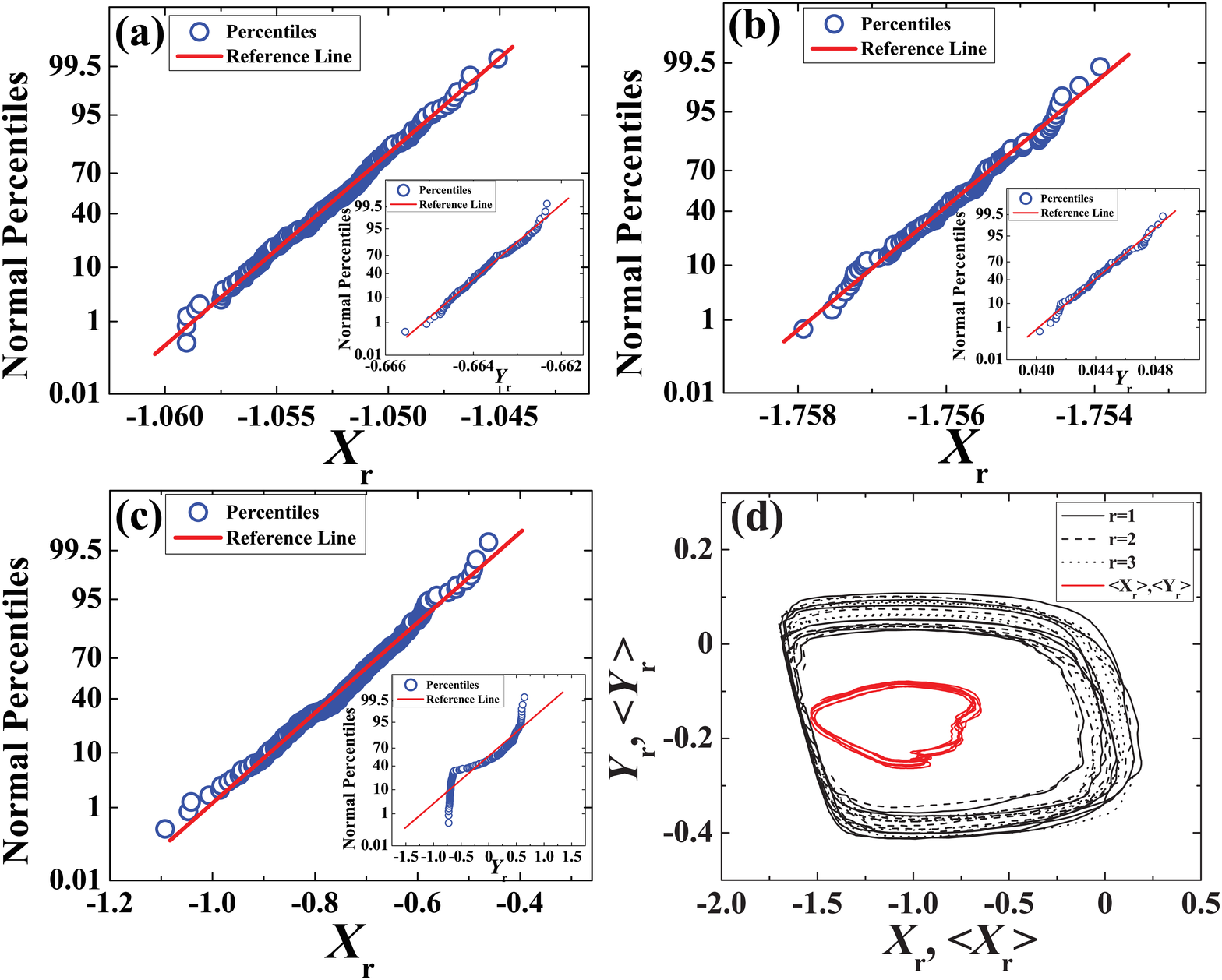}
\caption{(Color online) Examining the validity of QIA. Consistent with the theorem of large numbers, confirming the validity of GA for the global variables corroborates that QIA holds for the local variables. The main frames (insets) of (a), (b) and (c) refer to the graphic normality tests for the collective variables $X(t)$ ($Y(t)$), whereby the respective parameter sets correspond to those in Fig. \ref{Fig1}(a), (b) and (c). In (a) and (b) is demonstrated
that $X_r(t)$ and $Y_r(t)$ for different stochastic realizations are Gaussian distributed, whereas (c) indicates a substantial departure from the normal distribution in case of $Y_r(t)$. (d) illustrates the loss of qualitative analogy between the oscillations characterizing the individual realizations and the expectation for $(c,D,\tau)$ from (c). \vspace{-0.3cm}
\label{Fig3}}
\end{figure*}

We now turn to the analysis on the fulfilment of QIA. The intention here is just to briefly mention the two methods that may be used to verify the validity of QIA by examining the exact system, whereas the main point, lying in the ability to predict the failure of QIA solely by the dynamical features of the MF model is left for the next section. Considering the exact system, one may either take (i) an indirect approach, derived from a corollary of the QIA formulation, or (ii) the direct approach, based on the notion that approximate synchronization between the units may render them virtually independent. Since the more comprehensive discussion on the relation between different types of synchronization and the QIA is provided in subsection \ref{synchro}, we focus on the indirect approach (i). One first invokes the central limit theorem, by which for large, but finite $N$ holds that if the local variables are normally distributed for most $t$, so too are the collective variables. Hence, the validity of GA for $X(t)$ and $Y(t)$ should imply that the local variables are independent. The normality tests on $X(t)$ and $Y(t)$ are carried out analogously to those for $x_i(t)$ and $y_i(t)$. The main frames (insets) of Fig. \ref{Fig3}(a), (b) and (c) refer to graphic normality tests on the variable $X(t)$ ($Y(t)$) for the parameter sets exactly matching those in Fig. \ref{Fig1}(a), (b) and (c). Figures \ref{Fig3}(a) and (b) indicate the validity of GA for $X(t)$ and $Y(t)$ distributions, thereby suggesting that the QIA also applies. The positive result in Fig. \ref{Fig3}(b), which is associated with the oscillatory state, again draws on the main feature of the class II excitable systems. An interesting point regarding Fig. \ref{Fig3}(c) is that the distribution of $X_r(t_0+\delta t)$ over stochastic realizations conforms to, and the one for $Y_r(t_0+\delta t)$ sharply deviates from the Gaussian form. Such a violation of QIA is mostly found for intermediate $D$ and $\tau$. Figure \ref{Fig3}(d) further illustrates the loss of qualitative analogy between the oscillations for the individual realizations and the expectation, with the latter failing to preserve the relaxation character.

\section{Predicting the failure of QIA by the dynamics of the MF model} \label{bistab}

The aim in this section is to demonstrate how the failure of the QIA is indicated by the dynamics of the MF model.
To this end, we first present the results of the bifurcation analysis for the approximate system. Note that the analysis has not been carried out on the full system \eqref{eq6}, but rather on its counterpart obtained by retaining the equations for the first moments under the "adiabatic" approximation that the evolution of second moments is slow and can be cast as stationary. The main reason for this lies in the fact that the reduced system, unlike the original one, is analytically tractable.

\begin{figure}[t]
\centering
\includegraphics[scale=0.28]{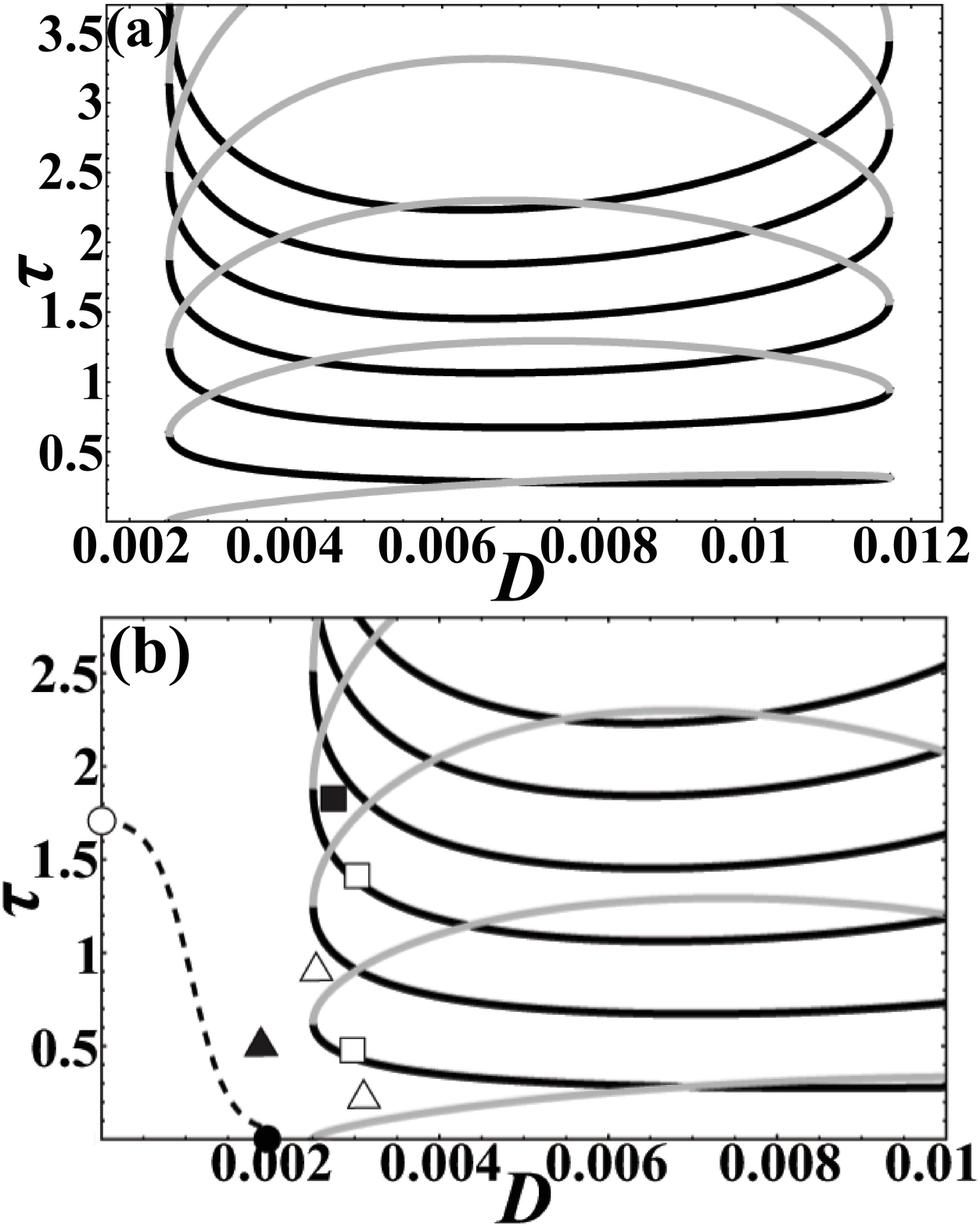}
\caption{(Color online) (a) First few branches $j=0,1,\dots,6$ of the Hopf bifurcation curves $\tau_{\pm}^j(D)$ for the MF model. (b) A close-up view of (a), but including the additional indication on the parameter values where the global fold-cycle bifurcations occur. Stability of equilibrium is influenced by a sequence of direct (supercritical) and inverse (subcritical) Hopf bifurcations, shown by the black and gray lines, respectively. In (b), the critical values $D_{fc}$ and $\tau_{fc}$ for the $D$- and $\tau$-controlled fold-cycle bifurcations are indicated by the solid and the open circle lying at $(D,\tau)=(D_{fc},0)$ and $(D,\tau)=(0,\tau_{fc})$. The dashed line approximately highlights the parameter values above which the dynamics of the MF model always involves a large cycle born via the global bifurcation. The bistable regimes emerging due to global bifurcations involve coexistence of FP and LC (instances indicated by triangles) or two LCs (instances indicated by the squares). For $D>D_H$, the existence of bistable regimes and their form depend on the complex interplay between the local and the global bifurcations. Coupling strength is fixed at $c=0.1$. \vspace{-0.4cm}
\label{Fig33}}
\end{figure}

The approximate model is found to display a sequence of supercritical and subcritical Hopf bifurcations, whereby the former (latter) result in creation of a stable (unstable) limit cycle. Recall that both types of Hopf bifurcation can further be cast as direct or inverse \cite{W00}, which refers to whether the fixed point unfolds on the unstable or the stable side, respectively. The final expression for the critical time delay in dependence of $c$ and $D$ reads \cite{derivation1,derivation2}:
\begin{align}
\tau_{\pm}^j&=[\arccos(-\kappa\epsilon/c)+2j\pi]/\omega_{\pm}, \,\,\text{if}\,\, \frac{-\omega_{\pm}^2+1/\epsilon}{c\omega_{\pm}/\epsilon}\geq0,\:\text{or}\nonumber\\
\tau_{\pm}^j&=[-\arccos(-\kappa\epsilon/c)+2(j+1)\pi]/\omega_{\pm}, \,\,\text{if}\,\, \frac{-\omega_{\pm}^2+1/\epsilon}{c\omega_{\pm}/\epsilon}<0, \label{eq5}
\end{align}
where the $+/-$ sign reflects the direct/inverse character of bifurcation, $j=0,1,2,\dots$ and
$\omega_{\pm}=\omega_{\pm}(c,D), \kappa=\kappa(c,D)$. It can be shown by a rather lengthy calculation that the direct (inverse) bifurcations are always supercritical (subcritical) \cite{derivation1}. The first few branches $j=0,1,\dots,6$ of the Hopf bifurcation curves $\tau_{\pm}^j(D)$ for the intermediate coupling strength $c=0.1$ are presented in Fig. \ref{Fig33}. In particular, Fig. \ref{Fig33}(a) is focused on the Hopf curves alone, whereas Fig. \ref{Fig33}(b) presents a zoom in of Fig. \ref{Fig33}(a), but also contains additional information, as explained below. Note that the presentation scheme in both figures is such that the curves coinciding with the direct (supercritical) Hopf bifurcations are indicated by the black lines, while those corresponding to inverse (subcritical) Hopf bifurcations are plotted by the gray lines.

Apart from the local bifurcations which affect the stability of equilibrium, the MF dynamics are influenced in a highly nontrivial fashion by the two global fold-cycle (tangent) bifurcations, one controlled by $D$ and the other by $\tau$. Note that the direct fold-cycle bifurcation gives rise to a stable large cycle and a saddle cycle. The point $(D,\tau)=(D_{fc},0)$ where the noise alone is sufficient to induce the global bifurcation is indicated by the solid circle in Fig. \ref{Fig33}. In an analogous fashion, the point $(D,\tau)=(0,\tau_{fc})$ where solely the delay gives rise to the global bifurcation is denoted by the open circle. The dashed line connecting the open and the solid circle approximately highlights the parameter values above which the dynamics of the MF model always involves a large cycle born via the global bifurcation. One should caution that in the parameter domains allowing for the local bifurcations, the existence of a large cycle \emph{per se} does not warrant multistability in the MF dynamics. Later on, it is shown that multistability in such domains depends on a complex interplay between the attractors and saddles resulting from the local and global bifurcations.

Due to global bifurcations, the MF model exhibits two types of bistable regimes, one involving the coexistence between the FP and the LC, and the other characterized by the coexistence of two LCs. In the former case, the LC corresponds to a large cycle born in the fold-cycle bifurcation. The latter scenario may be realized either by the coaction of the local (direct supercritical) Hopf bifurcation and the global fold-cycle bifurcation, which mainly occurs for $\tau<\tau_{fc}$, or the two cycles may both derive from the fold-cycle bifurcations ($\tau>\tau_{fc}$). In most cases, bistability emerges due to the action of noise, i.e. is facilitated by the $D$-controlled global bifurcation. Such regimes are referred to as the noise-induced bistability to distinguish them from the scenario involving the coexistence between the FP and the large cycle born in the $\tau$-controlled global bifurcation, which occurs for $\tau>\tau_{fc},D<D_{fc}$.

Our main point is that the noise-induced bistability in the dynamics of the MF model provides the necessary condition for the failure of QIA, and therefore the failure of MF approximation as a whole. In other words, the qualitative features of the dynamics displayed by the MF model can be used to predict in a self-consistent fashion the $(\tau,D)$ parameter domains where the QIA is bound to fail. Before explaining this point in more detail, we make a remark on why the noise-induced bistability is distinguished from the one owing solely to the $\tau$-controlled global bifurcation. Though the MF model makes no qualitative distinction between $D$ and $\tau$, which are both considered as equally valid bifurcation parameters, the exact system is naturally sensitive to the deterministic/stochastic character of the effects they generate. In this context, for $\tau>\tau_{fc}$ and sufficiently small noise, the oscillations displayed by the exact system retain their primarily deterministic character and as such satisfy the MF approximation trivially. Nonetheless, using the method described in section \ref{tests}, we have verified that the stochastic perturbation becomes large enough to compromise the validity of QIA for $D$ fairly close to $D_{fc}$, the noise intensity marking the onset of the $D$-controlled global bifurcation in the MF model.

\begin{figure}[t]
\centering
\includegraphics[scale=0.37]{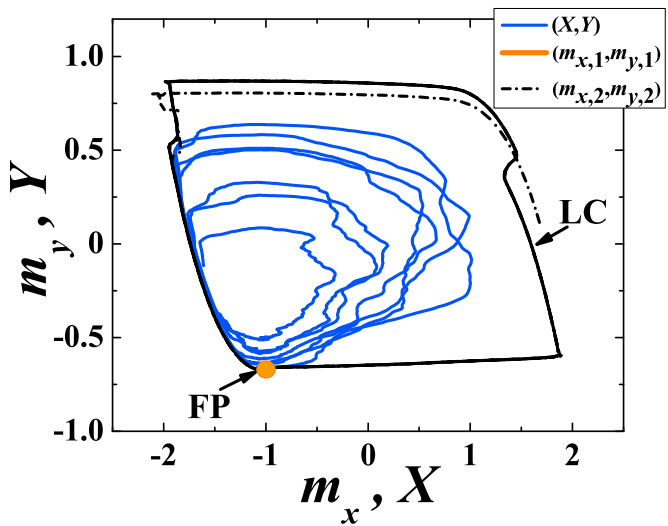}
\caption{(Color online) Bistability exhibited by the approximate model allows one to gain insight into the parameter domains where QIA breaks down. In the example provided, the MF dynamics $(m_x(t),m_y(t))$ shows coexistence of the FP, denoted by the orange (light gray) dot, and the LC (black dashed line) born via the global fold-cycle bifurcation. Influenced by noise, the typical orbit $(X(t),Y(t))$ of the exact system, displayed by the blue (dark gray) solid line, is found to fluctuate between the two attractors of the MF model. Failure of GA for the global variables may be considered an indirect evidence of the failure of QIA on the level of local variables. The data are obtained for the parameter set $(c,D,\tau)=(0.1,0.0029,0.3)$.
\vspace{-0.5cm}
\label{Fig4}}
\end{figure}

Next we show how the noise-induced bistability of the MF model is reflected in the dynamics of the exact system. First note that the illustrative examples of parameter values admitting bistability between the FP and the LC are indicated in Fig. \ref{Fig33} by the triangles, whereas the squares are reserved for the typical cases facilitating coexistence between the two LCs. In particular, we have singled out three instances related to bistability between the FP and the LC. The point denoted by the solid triangle ($\blacktriangle$) refers to the case bearing no influence from the local Hopf bifurcations, given that $D<D_H$, where $D_H\approx0.0025$ marks the onset of the Hopf bifurcations. Nevertheless, in the two examples indicated by the open triangles $(\vartriangle)$, such form of bistability occurs because the equilibrium is stabilized via the inverse subcritical Hopf bifurcation, whereby the unstable cycle born in the Hopf bifurcation acts like a threshold switching between the two stable solutions. Implementing the method introduced in section \ref{tests}, it has been verified that the QIA is violated in all the three described instances. For the case $(c,D,\tau)=(0.1,0.0029,0.3)$, we have illustrated the phase portraits corresponding to the two attractors of the MF model and the appropriate orbits for the collective variables of the exact system, see Fig. \ref{Fig4}. The rationale for the failure of QIA rests on the point that the mixed mode of the exact system may be interpreted as stochastic switching between the two attractors of the deterministic MF model. Naturally, the ensuing orbits are not normally distributed around the respective averages.

The analogous explanation also applies for the scenario where the MF model displays coexistence between the two LCs. In Fig. \ref{Fig33}, we have indicated three parameter domains supporting such form of bistability. In the cases denoted by the open squares $(\square)$, the large cycle from the global bifurcation coexists with the incipient cycle, emerging from the direct supercritical Hopf bifurcation. Nonetheless, the solid square $(\blacksquare)$ points to an instance where the two large cycles coexist, one of them created in the $D$-controlled, and the other in the $\tau$-controlled global bifurcation. It has been verified that the QIA breaks down in all of the stated instances.

One should note that crossing the Hopf bifurcation curves alone does not immediately imply the failure of the QIA. Nevertheless, due to interplay with the $D$-controlled global bifurcation, crossing the curves may become associated with the violation of QIA in two cases, one where the supercritical regime involves bistability of the FP and the LC (inverse subcritical Hopf bifurcation), and the other, which includes coexistence between two LCs (direct supercritical Hopf bifurcation). The occurrence of such cases is mostly confined to $\tau\lesssim\tau_{fc}$, because above $\tau_{fc}$ the MF dynamics is primarily influenced by the two global bifurcations.

\section{Miscellaneous topics} \label{misc}
\subsection{Fulfilment of MFAs and the statistics of the first return times} \label{return}

\begin{figure}[t]
\centering
\includegraphics[scale=0.185]{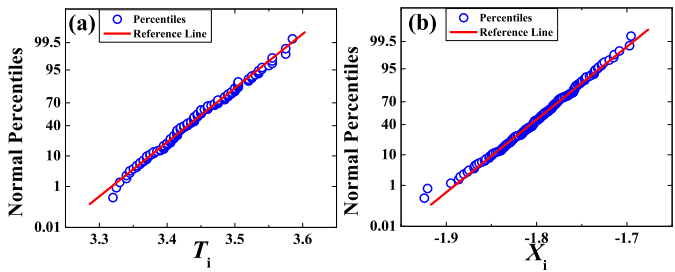}
\caption{(Color online) Characterizing the distribution of the return times and the return points for the macroscopic dynamics of the exact system. In (a) and (b) are displayed the graphic normality tests respectively indicating that the ISIs and the return points for $X(t)$ are Gaussian distributed. The data refer to the case of noise-led dynamics at $(c,D,\tau)=(0.1,0.0015,0)$, but one may arrive at qualitatively similar results for the delay-driven dynamics.
\vspace{-0.5cm}
\label{Fig5}}
\end{figure}

This subsection provides a discussion on some of the corollaries related to the fulfilment of the MFAs. Before elaborating on the relation between the synchronization properties and the fulfilment of QIA, we make two auxiliary notes qualifying more closely the terms "frequency" and "phase" used later on. The immediate aim is to show that the effective frequency and phase description of system dynamics may be appropriate if MFAs hold. Regarding frequency, we present the results on the distribution of ISIs for $X(t)$. Note that there are two types of collective modes, one where the ISIs are dominated by $T_0(D)$, which occurs for small and intermediate $D$ under very small $\tau$, and the other corresponding to the delay-led dynamics, which is typically seen for small and intermediate $D$ under large $\tau$. Either way, we have verified that ISIs are normally distributed for an arbitrary stochastic realization under long simulation times. In Fig. \ref{Fig5}(a), the normality test is provided for the more interesting case, showing persistence of Gaussian distribution for the noise-led dynamics under fairly large $D=0.0015$ at $\tau=0$. Since the analogous conclusion is readily reached for the delay-driven collective mode, one may state that the description of collective motion in terms of the average period (frequency) appears justified if MFAs apply.

A question that naturally arises is whether the fulfilment of MFAs implies that the distributions of the return points $P(X_r)$ and $P(Y_r)$ sampled at intervals equal to the average ISI of the macroscopic dynamics are also Gaussian. $P(X_r)$ and $P(Y_r)$ are calculated in two steps: one first lets the simulation run for the sufficiently long time to determine the average ISI for $X(t)$, and then carries on by collecting data on the return points for another very long time period. The first point $(X_0,Y_0)$ is chosen to lie on the refractory branch of the LC. In Fig. \ref{Fig5}(b) is displayed the graphic normality test for $P(X_r)$ along an arbitrary stochastic trajectory under the same parameter set as in Fig. \ref{Fig5}(a). The demonstrated normality of distribution indicates that, in statistical sense, the return points remain fairly close to the "average" LC. From the broader perspective, one may think of this result in the context of building an effective phase description for the collective motion \cite{piko1,piko2,piko3}.

Another point of interest is to verify whether the analogous conclusions hold for the local, rather than the global variables. Under the same parameter set as in Fig. \ref{Fig5}(a) and Fig. \ref{Fig5}(b), one can demonstrate for an arbitrary unit that the ISIs over a very long time series indeed conform to Gaussian distribution if the data from less than $10\%$ of realizations are discarded. However, the return points $P(x_r)$, sampled at $t_n=n*T_s$, where $T_s$ denotes the average ISI for the given unit, turn out not to be normally distributed. This is so because the Gaussian distribution for the local ISIs is comparably broader than the one for the global variables. Still, starting off from any point on the refractory or the spiking branch of slow motion, the successive return points, recorded at $T_s$ long intervals, always fall on the "right" branch, determined by the location of the initial point. Therefore, the above results suggest an interesting point that if the MFAs are satisfied, the use of terms frequency and phase is more appropriate to describe the dynamics of the global, than the local variables.

\vspace{-0.5cm}
\subsection{Fulfilment of QIA and synchronization} \label{synchro}

Having gained insight into the competition between the noise-led and the delay-driven dynamics, as well as the statistical features providing the context for the effective use of terms frequency and phase, we proceed with the analysis of the relation between the synchronization of the individual units and the fulfillment of QIA. To begin with, one notes that for being stochastic and excitable in nature, the units cannot exhibit complete synchronization. However, the discussion above suggests that it is reasonable to speak of approximate frequency (FS) and phase synchronization (PS) in conditional terms, viz. if MFAs are satisfied. Presence or absence of these forms of synchronization may give rise to three types of collective states: (i) coherent states where single units display both the approximate FS and PS, (ii) states that exhibit FS, but lack PS and (iii) collective states where approximate FS is not established. One may infer the relation between synchronization and QIA by examining the linear interaction terms of the form $c*(x_i(t-\tau)-x_j(t))$. If there is approximate lag-synchronization, the latter become very small, which leaves the neurons virtually independent. Therefore, by identifying conditions under which the approximate lag-synchronization is achieved, one effectively looks for the parameter domains where QIA applies.

\begin{figure*}[t]
\centering
\includegraphics[scale=0.295]{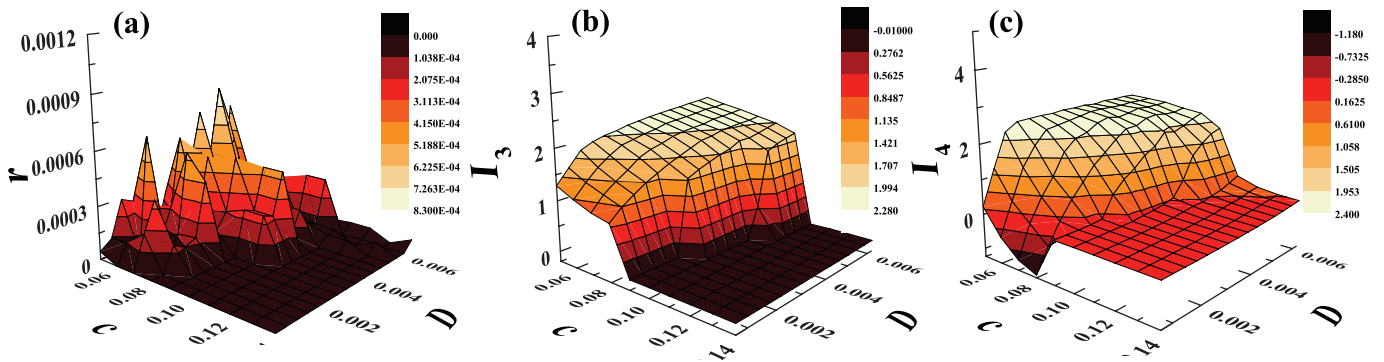}
\caption{(Color online) Focus on identifying the parameter domains that admit frequency and phase synchronization between the single units, which effectively provides an indication of where the QIA holds. (a) shows $r(c,D)$ for the delay-driven dynamics at $\tau=2.7$. In (b) and (c) are illustrated the corresponding $\overline{I_3}(c,D)$ and $\overline{I_4}(c,D)$ distributions for $\tau=2.7$, respectively. Comparing (a), (b) and (c), note the overlap between the regions displaying near-zero values for the $r$ ratio, $\overline{I_3}$ and $\overline{I_4}$. The analogous result can be obtained for the noise-driven dynamics at $\tau=0$.
\vspace{-0.5cm}
\label{Fig6}}
\end{figure*}

We have established that there exist only two scenarios for the approximate lag-synchronization, both of which amount to cases of approximate FS and PS. The interaction terms may substantially reduce either (i) for noise-led dynamics at $\tau\simeq0$, or (ii) for delay-driven dynamics at very large $\tau\sim T_0(D)$. A way to characterize the approximate FS for the given parameter set is to calculate the ratio $r=\Delta T/\langle \overline{T_i}\rangle$, where $\Delta T=max|\overline{T_i}-\overline{T_j}|$ is the maximal difference between the time-averaged ISIs $\overline{T_i}$ of individual units, whereas $\langle\overline{T_i}\rangle$ denotes the population average $\langle\overline{T_i}\rangle=N^{-1}\sum\limits_{i=1}^{N}\overline{T_i}$. The smaller $r$ becomes, the better FS between the units is achieved. The results for $r(c,D)$ plotted in Fig. \ref{Fig6}(a) refer to the (ii) case at $\tau=2.7$. We have verified that setting $\tau=0$, which corresponds to case (i), yields qualitatively similar results. As the main point, note a very large domain where $r$ is small, which indicates the approximate FS. Expectedly, for small $c$ and large $D$, $r$ is seen to rise sharply, implying that FS is lost.

The drawback of the method above is that one cannot distinguish whether approximate FS is or is not accompanied by PS. To do so, we consider the time-averaged third and fourth order moments of the local potentials $P(x_i(t))$ for the given parameter set, taking the average over a very long stochastic realization. Note that if the ergodic hypothesis applied, such an average would equal the one over an ensemble of realizations. Nevertheless, whether this holds or not is of marginal significance because the results below are not intended to be rigorous, but should rather provide an illustration on the link between PS and the fulfilment of QIA. Therefore, the discussion on the asymptotic distributions here is independent and should by no means be confound with the results from section \ref{tests}, which only concern averaging over an ensemble of stochastic realizations.

As for the moments, the third-order average moment is defined by $\overline{I_3}=(1/T)\sum\limits_{t=1}^{T}I_3(t)$, where $I_3(t)=\sum\limits_{x_i}(x_i-X(t))^3P(x_i(t))$. The analogous relation holds for $\overline{I_4}$. $P(x_i(t))$ is obtained by dividing the range of possible $x_i$ values into $110$ bins $[x,x+\delta x]$, whereby one records the fraction of units whose potential falls within the given bin. If there is an approximate FS and PS, one expects $x_i$ for most $t$ to be Gaussian distributed around the mean $X(t)$. Then, both $\overline{I_3}$ and $\overline{I_4}$ should lie close to zero. If there is approximate FS, but PS is lacking, $\overline{I_3}\approx0$ should hold, whereas $\overline{I_4}$ should substantially depart from zero. Finally, if there is no approximate FS, both $\overline{I_3}$ and $\overline{I_4}$ are supposed to lie away from zero. Results on $\overline{I_3}$ and $\overline{I_4}$ at $\tau=2.7$ for a wide range of $(c,D)$ values, cf. Fig. \ref{Fig6}(b) and \ref{Fig6}(c), suggest that domains with approximate FS closely match those with PS. Note the overlap between the areas with the smallest $r$, $\overline{I_3}\approx0$ and $\overline{I_4}\approx0$ in Fig. \ref{Fig6}(a), \ref{Fig6}(b) and \ref{Fig6}(c), where QIA should hold.

\section{Conclusion} \label{conc}

The reduction of computational demand and the possibility of describing the stochastic stability and the stochastic bifurcations can be cast as general reasons for introducing the MF approximate model for an arbitrary set of SDDEs. Given the apparent relevance of the MF method, an issue of considerable importance is to be able to determine the domains where such an approach may provide accurate qualitative predictions. The approximations behind the MF model are often considered in a simplistic fashion, as if they were completely independent on the class of systems which the model under study belongs to. Such a view results by invoking the (stereo)typical requirements for small noise intensity and weak couplings as the  main conditions for the validity of the MF model.

In the present paper, the issue of the MF approximations and their validity is highlighted by taking the example of a system of delay-coupled noisy type II excitable units, represented by the generic Fitzhugh-Nagumo model.  What we actually show is that, though they contain certain commonly stated elements, the MFAs relevant for the given system also include ingredients that should be precisely adapted to its essential dynamical properties. In particular, the inherent features of class II excitable systems, such as relaxation character of oscillations, have been explicitly incorporated into the definitions of the two MFAs we introduced. This point is particularly apparent in the definition of the GA, and is further reflected in the fashion in which the validity of both the GA and the QIA has been verified.

It is found that the requirements for the joint validity of GA and QIA may be expressed in terms of a single qualitative statement, by which the two apply if the local and global dynamics exhibit a unique attractor of the same type, either a FP or a LC, provided that $D$ is not overly large. Of the two generic scenarios, the one involving the stochastically stable FP is fairly trivial, whereas the one associated with the LC is more intricate and makes apparent the need for introducing the refined MFAs considered in the paper.

Focusing on each of the approximations independently, it is shown that validity of GA cannot be explicitly tied to certain parameter domains, but rather comes down to a qualitative requirement for not too large a noise intensity. This is the main corollary of the actual statement on the validity of GA, by which GA is satisfied if the qualitative similarity between the individual realizations and the appropriate expectations is maintained for the given parameter set. For the oscillatory state, the notion of qualitative similarity effectively refers to the point that the expectations preserve the relaxation character of oscillations. In this context, we have attempted to provide some quantitative measure on validity of GA by determining the variation of the $N_{out}/N_r$ ratio with $D$, see section \ref{tests}.

Nonetheless, our main conclusion regarding the validity of MF approximation is associated with the fulfillment of QIA. What we have demonstrated is that the failure of QIA can explicitly be related to the noise-induced bistability of the MF model. Such bistable regimes, involving either the coexistence between the FP and the LC or the two LCs, are influenced by the global fold-cycle (tangent) bifurcation controlled via the noise intensity parameter. In this fashion, the $(\tau, D)$ parameter domains where the MF approximation is bound to fail are identified with the domains admitting noise-induced bistability for the MF model's dynamics. In other words, the noise-induced bistability of the MF model provides the necessary condition for the failure of the QIA, and thus the MF approximation. Note that such parameter domains do not exhaust all the cases where the MF approximation fails, because the breakdown may also be caused by the violation of GA.

As for the relationship between the Hopf bifurcation curves determined for the MF model and the failure of MF approximation, we stress that crossing the curves itself does not imply the failure. It has already been pointed out that the latter would mean that the MF model could never account for the collective oscillatory states, which is not true. Though the asymptotic distribution for the collective variables in the exact system indeed loses the Gaussian property if the curve corresponds to the stochastic Hopf bifurcation, this fact alone has no bearing on the MF approximations we introduced. However, in the interplay with the $D$-controlled global bifurcation, crossing the Hopf bifurcation curves may involve the onset of two different bistable regimes in the MF model, and as such, may contribute to the violation of the QIA, and thereby the MF approximation as a whole. It is reasonable to expect that the scope of the conclusion on the relationship between the noise-induced bistability of the MF model and the failure of MF approximation may likely be extended to a broader range of systems, since it draws only on the qualitative properties of the system dynamics. For the future research, it would be interesting to examine the refined MFAs and their validity by carrying out the analysis similar to ours in case of the MF models derived for systems exhibiting complex multi-scale oscillations, such as bursting, when subjected to noise and coupling delays.

\begin{acknowledgments}
This work was supported in part by the Ministry of Education and
Science of the Republic of Serbia, under project No. $171017$ and
No. $171015$.
\end{acknowledgments}

\end{document}